
\magnification\magstep1
\baselineskip=16pt
\def\no{\noindent}
\no
\centerline{\bf Spontaneous magnetization in the Disorder dominated phase}
\centerline{\bf of the Two-dimensional}
\centerline{\bf Random Bond Ising Model}
\vskip 0.5in
\centerline{short title: Magnetization in the 2D Random Bond
Ising Model}
\vskip 0.3in
\centerline{D.Braak}
\vskip 0.3in
\centerline{Institut f\"ur Theorie der Kondensierten Materie, Universit\"at
Karlsruhe,}
\centerline{Physikhochhaus, D-7500 Karlsruhe, Germany}
\vskip 0.3in
\no
Abstract:\par
The selfconsistent approach to the two-dimensional
Ising model with quenched random bonds is extended to the
full lattice theory described by four real fermions. A calculation
of the averaged spin-spin correlation function for large separation of the
spins in the disorder dominated phase indicates an
exponential decay of this quantity and therefore a vanishing spontaneous
magnetization. The corresponding correlation length is proportional to
$1/\eta^2$, where $\eta$ denotes the order parameter of the new phase
introduced by Ziegler.
\vskip 0.8in
\no
\vfill
\eject
\no
{\bf 0. Introduction}
\bigskip
\no
The 2D-Ising model [1], defined on a square lattice with
quenched disorder in the ferromagnetic bonds is defined via the Hamiltonian
$$
H=-{1\over{k_BT}}\sum_{<i,j>}J(i,j)S_iS_j.
$$
The bond strengths are ferromagnetic random variables $0<J(i,j)$ and
$\overline{J(i,j)} = J_0$. $<i,j>$ is a pair of next neighbor sites.

\no
Whereas it is well confirmed that this system undergoes a phase
transition to
a ferromagnetic ordered state at a lowered critical temperature
compared with the pure system, the nature of the transition is not
completely understood because of the non-applicability of the Harris
criterion
and several conflicting analytical results [2,3,4].
The only exactly solvable
model for a disordered ferromagnet is the McCoy-Wu model [5],
where the disorder
is essentially one-dimensional. For the model with isotropic disorder the
famous analysis by V.Dotsenko and Vl.S.Dotsenko [4] mapped the system
onto the $N=0$-Gross-Neveu model to get the averaged thermodynamic
quantities. This is possible because of the free fermion representation
of the pure model. The authors used the replica trick and a continuum
field theory, which was analyzed with the momentum space renormalization
group (RG-) technique. This model is asymptotically free in the infrared,
the coupling $g$ ($\simeq$ disorder strength) is marginal irrelevant,
the phase diagram is not changed and the specific heat diverges at the
critical point but more slowly than in the pure model.($\simeq \log\log
(|T-T_c|^{-1})$ versus $\log (|T-T_c|^{-1})$).
The averaged spin correlation
function $\overline{<S_0S_n>}$ shows a drastic change, namely slow decay:
$\overline{<S_0S_n>}\simeq \exp -{1\over{g}}(\log\log n)^2$. The disorder
seems to enhance the correlations instead of destroying them, the
critical exponent goes from 1/4 to zero.

\no
An alternative approach uses the bosonized version of the pure model,
which makes the spin operators local [6,7,8].
The disorder is treated with the
same method as in [4]. The results confirm the behavior of the specific
heat but predict only logarithmic corrections to the correlation exponent.
An equivalent way is to take advantage of the conformal invariance of the
pure model [9]. All these methods bear two difficulties: first, the replica
trick is used and second, they start from the continuum limit associated
with the supposed critical point. Now, there exist an alternative to
the replica trick through a supersymmetric version of the effective model
[3,10]. This is important because the replica trick is questionable in our
case [11,12]. [10] performs a RG-analysis of this model and [3] calculates
the saddle point structure of the theory after a transformation to
composite operators (Q-matrices). Both [3] and [10] show finite specific
heat in the critical region. Whereas the RG-treatment in [10] lacks the
correct incorporation of the additive renormalization crucial to
extensive quantities [13], [3] shows that in the critical region a
new saddle point becomes stable and governs the thermodynamic behavior of
the effective model. This saddle point is accompagnied by spontaneous
symmetry breaking and a new phase between the ferro- and paramagnetic
one. The corresponding order parameter stems from a regularization term
needed for the boson integration in the supersymmetric theory [3,10].
The non-vanishing of the order parameter can be proven rigorously [14,15],
but there exist two regularizations corresponding to different physical
situations, one to the Random bond Ising model, the other to a system
of polymer chains [12].

\no
All investigations mentioned so far use a large scale approximation where
two of four lattice-fermion degrees of freedom are
ignored. The RG-approaches
even need the continuum limit. The difficulty of the continuum limit
is connected with the fact that the renormalization procedure
 and the bosonization are not interchangeable. This is the ``technical''
reason for the discrepancy of the findings in [4], respective [5,6,7]
regarding the spin correlation function [16].

\no
 Therefore it seems worthwile
to extract as much information on the magnetic correlations as possible
without large scale approximation. This is the aim of the present article.
Moreover, the technique used avoids both the
replica trick and the supersymmetric theory, staying close to the original
model by direct average of the Greensfunction in the framework of a
$1/N$ expansion [2]. The parameter $N$ serves as a bookkeeping device
to derive a selfconsistent theory which contains the tadpole structure
completely (section 1). The diagrammatic expansion builds a bridge between
the Q-matrix approach and the RG-calculations by comparison of the class
of Feynman diagrams which are accounted for. The tadpoles are usually
omitted in field theoretical investigations because they can be treated by
normal ordering. But this may be dangerous if one does not know the
vacuum structure, i.e. the phase diagram. With the full 4-fermion lattice
theory one can identify the diagrams which yield a contribution second
order in the coupling to an ``exceptional'' mass term (see section 1),
which turns out to be equivalent to the regularization term in [3].
Eventually, this term (the order parameter of the new phase)
cannot be viewed as an artefact of the supersymmetric theory. But the main
reason to consider the full theory is the possibility to calculate
magnetic correlations which are the subject of the many computer simulations
performed on the model [17,18,19]. These show very good agreement with the
thermodynamic predictions of [3](see [2]; an experimental result for the
specific heat is also available [20])

\no
The disorder average of the square of the spontaneous magnetization in the
new phase is now calculated in section 2.
This is done by determination of  the averaged
spin correlation function
for large separation of the spins (section 2).
This quantity is related to the averaged square of the extensive spontaneous
magnetization, i.e.
$\overline{{\cal{M}}^2}=\overline{(\sum_{i}\langle S_i\rangle)^2}$,
which is translationally invariant before the disorder average,
by means of a Griffith inequality [21] and the cluster property [21,22]
together with the hypothesis of self-averaging of ${\cal{M}}$.
(The  spin correlation function itself, whose first moment is calculated in
this article needs not to be self-averaging [9]. But exponential
decay of the magnetic correlations $\overline{\langle S_iS_j\rangle}$
is sufficient for the conclusion, that ${\cal{M}}$ vanishes in a
{\it{fixed}} sample in the thermodynamic limit.)
Even if the cluster property is questionable [22], our
result (${\cal{M}}=0$) is
not affected by this. However, $\overline{{\cal{M}}^2}$
 should
not be confused with the Edwards-Anderson order parameter
for spin glasses [22]:
 $\overline{\langle S_i\rangle ^2}$, a local quantity,
which becomes translationally invariant
after the average over disorder. Therefore, the present methods do not
allow for an estimate of this important quantity in the new phase.
We use Gaussian disorder to simplify the calculations although a bounded
distribution of the bond strenght is necessary to keep all the bonds
ferromagnetic. Nevertheless, the only relevant cummulant in the $N=\infty$
limit is the second one.
 In [2] it was shown that the higher
cummulants appear only in higher order terms in the
$1/N$-expansion.
\vfill\eject

\def\r{{\bf{r}}}
\def\p{{\bf{p}}}
\def\a{\alpha}
\def\b{\beta}
\def\1{{\bf{1}}}
\def\ph{\phantom{0}}
\def\g{\gamma}
\def\d{\delta}
\def\e{\epsilon}
\no
{\bf 1. The selfconsistency equation.}

\no
The N-fold replicated partition function
for a specific configuration of
disorder is
$$
Z_{DN}=\int {\cal{D}} \xi \exp {H_{D}}
$$
with the euclidean action (Hamiltonian) $^{1,2}$
$$
H_{D}=\sum_{\a,\b}^{N}\sum_{\r}
\xi_{1}^{\a}(\r)\xi_{2}^{\a}(\r)+\xi_{3}^{\a}(\r)\xi_{4}^{\a}(\r)
+\xi_{1}^{\a}(\r)\xi_{4}^{\a}(\r)+\xi_{1}^{\a}(\r)\xi_{3}^{\a}(\r)
+\xi_{3}^{\a}(\r)\xi_{2}^{\a}(\r)+\xi_{2}^{\a}(\r)\xi_{4}^{\a}(\r)
$$
$$
+t_{0}\xi_{1}^{\a}(\r)\xi_{2}^{\a}(\r +e_{y})
+t_{0}\xi_{3}^{\a}(\r)\xi_{4}^{\a}(\r +e_{x}) \eqno (1)
$$
$$
+\delta t_{y}^{\a \b}(\r)\xi_{1}^{\a}(\r)\xi_{2}^{\b}(\r +e_{y})
+\delta t_{x}^{\a \b}(\r)\xi_{3}^{\a}(\r)\xi_{4}^{\b}(\r +e_{x})
.$$
The $\xi_{i}^{\a}$ are real Grassmann fields defined at the points $\r$ of a
two dimensional square lattice $\Lambda =Z\times Z$,
$e_{x,y}$ are the unit vectors in the two
directions of the lattice. $t_{0}=\tanh J_{0}/kT$ measures the average bond
strength. The disorder variables $\delta t_{x,y}^{\a \b}$ are statistically
independent and Gaussian distributed with mean zero and variance
($i,j=x,y$)
$$
\overline{\delta t_{i}^{\a\g}(\r)\delta t_{j}^{\b\d}(\r ')}
={g\over{N}}\delta _{\r ,\r '}\delta _{i,j} [\delta _{\a \b}\d_{\g\d}
+\d_{\a\d}\d_{\b\g}].
\eqno (2)$$
For a given configuration $D$ of disorder one expands the Green's function
for a fixed index $\a$:
$$
G^{\a,D}_{i,j}(\r,\r')=\langle \xi^{\a}_{i}(\r)\xi^{\a}_{j}(\r')\rangle
_{A_{D}}
\eqno(3)
$$
With
$$
G_{i,j}^{\a,0}(\r,\r')=\langle \xi^{\a}_{i}(\r)\xi^{\a}_{j}(\r')\rangle
_{\delta t=0}
\eqno (4)
$$
one gets
$$
G^{\a,D}_{i,j}(\r,\r')=G^{\a,0}_{i,j}+
$$
$$
\langle \xi^{\a}_{i}(\r)
\sum_{m=1}^{\infty}\sum_{\a_{k},\b_{k}}^{N}\sum_{\r_{1}\ldots \r_{m}}
\prod^{m}[\delta t_{y}^{\a_{k},\b_{k}}(\r_{k})\xi_{1}^{\a_{k}}(\r_{k})
\xi_{2}^{\b_{k}}(\r_{k}+e_{y})
\eqno (5)
$$
$$
+
\delta t_{x}^{\a_{k},\b_{k}}\xi_{3}^{\a_{k}}(\r_{k})\xi_{4}^{\b_{k}}
(\r_{k}+e_{x})
]\xi^{\a}_{j}(\r')\rangle_{\delta t=0}.
$$
Applying the fermionic Wick theorem and averaging over the disorder fields
to get the averaged Green's function $G'(\r,\r')=\overline{G^{\a}(\r,\r')}$
leads to a closed loop of $G^{0}$-propagators with selfcontacts.
 As was shown previously$^{2}$
the only terms contributing to order $N^{0}$ have the tadpole structure.
There are essentially four of them (i.e. eight for both axes),
 depicted in fig.1 and fig.2. The left
graph of fig.1, for example, corresponds to
$$
{g\over{N}}
\sum_{\b}^{N}\sum_{\r_{1}}G^{\a,0}_{i,1}(\r ,\r _{1})
G^{\b,0}_{2,1}(\r _{1}+e_{y},\r _{1})
G^{\a,0}_{2,j}(\r _{1}+e_{y},\r ')
\eqno (6)
$$
Therefore one deduces a (nonlinear) selfconsistency equation including all
tadpoles. Dropping the  replica indices it reads
$$
G'_{k,l}(\r ,\r ')=G^{0}_{k,l}(\r ,\r ')+
$$
$$
g\sum_{\r _{1}}\sum_{i}^{4}\sum_{j=i,\hat{i}}(-1)^{\delta _{\hat{i},j}}
G^{0}_{k,i}(\r ,\r _{1})
G'_{\hat{i},j}(\r _{1}+
e(\hat{i}),\r _{1}+e(j))G'_{\hat{j},l}(\r _{1}+ e(\hat{j}),\r ')
\eqno (7)
$$
with
$$
\hat{1}=2,\hat{2}=1,\hat{3}=4,\hat{4}=3
$$
and
$$
e(1)=-e(2)=-e_{y}
$$
$$
e(3)=-e(4)=-e_{x}.
$$
The sign factor in (7) comes from an exchange of Grassmann variables in
case of diagonal propagators in the loop (see fig2.).
 Equation (7)
determines the ``selfenergy'' of the
effective Hamiltonian. In momentum space
one gets with $H'=G'^{-1}$ and $H_{0}=(G^0)^{-1}$
$$
H_{0}(\p)=H'(\p)+gC(\p).
\eqno (8)
$$
$C$ is a $4 \times 4$ matrix
$$
C=\pmatrix{c_{0y}&e^{ip_{y}}c_{y}&0&0\cr
-e^{-ip_{y}}c_{y}^{*}&c_{0y}&0&0\cr
0&0&c_{0x}&e^{ip_{x}}c_{x}\cr
0&0&-e^{-ip_{x}}c_{x}^*&c_{0x}\cr}
\eqno (9)
$$
with
$$
c_{0y}=-\int_{-\pi}^{\pi} {dp_{x}dp_{y}\over {2\pi}}G'^{11}(\p)
$$
$$
c_y=\int_{-\pi}^{\pi} {dp_xdp_y\over {2\pi}}e^{ip_{y}}G'^{21}(\p)
\eqno (10)
$$
$$
c_x=\int_{-\pi}^{\pi} {dp_xdp_y\over {2\pi}}e^{ip_x}G'_{43}(\p)
$$
Now for $x/y$-symmetric disorder, $c_{0x}=c_{0y}$ and the kernel of
the Hamiltonian of
the pure system reads$^1$
$$
H_0(t_0,\p)={1\over {2}}\pmatrix{0&a&1&1\cr
-a^*&0&-1&1\cr
-1&1&0&b\cr
-1&-1&-b^*&0\cr}
\eqno (11)
$$
with $a=1-t_0e^{ip_y}$ and $b=1-t_0e^{ip_x}$. $t_0=\tanh (1/k_BT)$.
Therefore, the most general ansatz compatible with the selfconsistency
equation is
$$
H'(t,\p)= H_0(t,\p)+{\eta\over 2} {\bf{1}}
\eqno (12)
$$
where ${\bf{1}}$ is the identity matrix. A nonvanishing $\eta \in R$
prevents $H'$ from getting  eigenvalues $=0$ for any value of $\p$ and the
divergency of the specific heat in the pure model vanishes.
In a large scale approximation
one recovers just the ``externally regularized`` model of [3], where
the possibility of the $\eta$-term was assumed for different reasons.
In fact, a saddle point calculation led to a $\eta \neq 0$ in a
narrow region around the (shifted) critical temperature which corresponds
to $t=t_c=\sqrt{2}-1$.

\no
In the present consideration of the full lattice theory, the $\eta$-term
is connected with the diagonal propagators of fig.2. These vanish if
$\eta =0$ because the pure Greensfunction $G^0_{ii}(\r ,\r )=0$. This
is clear as we started from a {\it{Majorana}}-field theory with real
fermions. The $\eta$-term is forbidden within such a field theory.
However, the average over disorder gives nonzero contributions to
the diagonal propagators second order in the coupling $g$. But the
selfconsistency condition allows for $\eta \neq 0$ only in a neighborhood
of the critical temperature as in [3]. Eventually the model must correspond
to an effectice {\it{Dirac}}-field theory in this region which allows for
the diagonal entries in $H'$. The  breaking of the
discrete symmetry in [3] can be understood as
``spontaneous charge generation'' in going from real to complex fermions.

\no
In our approach the condition for a nonvanishing $\eta$ deduced from (8)
is
$$
0=\eta {\Big (}1-g\int d^2\p {\eta ^2+2+bb^*\over {det(\eta,t)}}{\Big )}
\eqno (13)
$$
with
$$
 det(\eta ,t)=\det(2H'(\eta,t))=\eta ^4+4\eta ^2+\eta ^2(|a|^2+|b|^2)
-4Re(a)Re(b)+|a|^2|b|^2+4
\eqno (14)
$$
This expression is minimal for $t=t_c$, therefore it exists a region
around the point $t=t_c$ determined by
$$
1=g\int d^2\p {\eta ^2+bb^*+2\over {det(\eta,t)}}
\eqno (15)
$$
where $\eta\neq 0$.
This equation, together with
$$
t_0=t+g\int d ^2\p {{e^{ip_y}[a^*(|b|^2+\eta ^2)-2Re(b)]}\over{det(\eta,t)}}
\eqno (16)
$$
allows to calculate $\eta(t_0,g)$ and $t(t_0,g)$ in this region.
The phase boundaries to the ``outer phase'' with $\eta =0$ are given
by
$$
1=g\int d^2\p {{bb^*+2}\over{det(\eta =0,t(t_{0}^{\pm},g))}}.
\eqno (17)
$$
Equation (16) leads for $\eta = 0$ to
$$
t_0=t+g{\Big [}{t\over {1+t^2}}+{{3t^2-1}\over{4t(1-t^2)}}{\Big ]}
F({1\over2},
{1\over2};1;k^2)-g{{3t^2-1}\over{4t(1-t^2)}}
\eqno (18)
$$
with $F(k^2)$ the hypergeometric function and $k=4t(1-t^2)/(1+t^2)^2$.
Fig.3 shows the effective $t$ as function of $t_0$ and the two values
$t_-$ and $t_+$. In the next chapter it is shown that the spontaneous
magnetization vanishes in the new phase characterized by $\eta \neq 0$.
Therefore, $t_+$ has to be viewed as the transition point from the
ferromagnetic to the disordered phase. This leads to a lowered critical
temperature, as expected.
\vfill
\eject
\no
{\bf 2. The spontaneous magnetization in the disordered phase.}

\no
To get the spontaneous magnetization, we calculate the
averaged spin-spin
correlation function  along the
x-axis $\overline{S(n)}=\overline{\langle \sigma (0)\sigma (n)\rangle}$
for large values of the spin distance $n$ measured in units of the
lattice spacing. This quantity is connected to the magnetization through
the relation$^1$
$$
\overline{{\cal{M}}^2}=\lim _{n\rightarrow \infty} \overline{S(n)}
\eqno (19)
$$
The following discussion has two limitations:

\no
1. We use the effective Hamiltonian derived in section 1., the inverse of
the disorder averaged Greenfunction for the fermion operators.
 This leads to the omission of
certain contributions to the averaged correlation function of the spin
operators which are nonlocal in the fermions$^1$. Nevertheless, this
is correct within the $N=\infty $ approximation because the omitted terms
are of higher orders in $1/N$. For a
specific configuration $D$ of disorder, $S_D(n)$
can be written as ratio of two
partition functions$^1$
$$
S_D(n)=\prod_{i=0}^{n-1}t_x(i,0){Z'_D\over{Z_D}}
\eqno (20)
$$
where in $Z'_D$ the bonds along the
x-axis between the sites $(0,0)$ and $(n,0)$
are modified from $t_x(i,0)$ to $t_x^{-1}(i,0)$. This amounts to
replace $\xi H_D\xi$ in the
functional integral by $\xi H_D\xi +{1\over2}\xi Q\xi$ and
$$
{1\over2}\xi Q\xi ={{1-t_x^2(i,0)}\over{t_x(i,0)}}\sum_{i=0}^{n-1}
\xi_3(i,0)\xi_4(i+1,0)
\eqno (21)
$$
 Now $\langle\exp {1\over2}\xi Q\xi\rangle_{Z_D}$ is
expanded$^4$
in a series of closed loops of products of fermionic propagators atached to
the line between $(0,0)$ and $(n,0)$ (see Fig.4). Averaging
 now these quantities
leads to contributions of the forms depicted in Figs.5, 6 and 7.
If we replace in
formulas (20) and (21) $t_x(i,0)$ by $t$ and $H_D$ by $H'$ as derived in
sect.1, the graphs of Fig.5 and 6. are counted but not those of
 Fig.7. However, only
the first two are of order $N^0$ in the $1/N$ expansion.
 This corresponds to the
fact that the level of a renormalization group treatment is not reached in
this consideration on "mean field level", which is suitable
 to handle the nontrivial phase diagram
of the problem. Nevertheless, the graphs of fig.7 can be factorized from the
contributions which are accounted for in the expansion of the averaged
correlation function. Therefore they
cannot affect the result of the following
calculation, i.e. they cannot lead
to a nonvanishing spontaneous magnetization
in the new phase.

\no
2. The results are only valid in a small region around $t_+$, i.e. for small
$\eta$ and welldefined $G_0(t(t_0, g))$. In addition we use Szeg\" o's
lemma$^1$ to evaluate the spin correlation
function of the associated ``pure''
system, i.e. with the modified coupling $t$ instead of $t_0$. This gives
results only in the limit $n\rightarrow \infty$. Nevertheless our result
is applicable for large but finite $n$, because in the vicinity of $t_+$
the strong $n$-dependence of $\overline S(n)$ in the $\eta \neq 0$
region is multiplied by an almost constant factor coming from the
nonvanishing spontaneous magnetization for $t<t_c$.
Moreover, the width of the region of validity depends only on $t_+$, not
on $n$. The approximations made in the appendix affect the analytical form
of the dependence of the correlation
lenght on $\eta$ only (see equation (37)).
They have no influence on the qualitative behavior of the averaged
correlation function. Close to $t_+$ our result is correct at least in the
sense of the ``scaling limit'' results obtained in the pure model [1].

\no
Following [1] we write for the square of the disorder averaged spin
correlation function
$$
{\overline{S(n)}\ }^2=t^{2n}\det{\Big (}\1 -
{(1-t^2)\over t}G'(t,\eta )Q'{\Big )}.
\eqno (22)
$$
$Q'$ projects on the (3,4)-(i.e. horizontal) sector of the four fermions and
is off-diagonal in the position space index.
$$
Q'_{i,j}(x_1,y_1;x_2,y_2)=
\delta_{i,3}\delta_{j,4}\delta_{x_2,x_1+1}\chi(x_1)
\delta_{y_1,0}\delta_{y_2,0}
-\delta_{i,4}\delta_{j,3}\delta_{x_1,x_2+1}\chi(x_2)
\delta_{y_1,0}\delta_{y_2,0}
\eqno (23)
$$
$\chi(x)$ means the characteristic function on the intervall $[0,n]$.
Eventually, $Q'$ projects $G'$ on a $2(n+1)\times 2(n+1)$ dimensional
subspace.
Therefore, we have to evaluate the determinant of the $2(n+1)\times 2(n+1)$
matrix
$$
\tilde{t\1}+(1-t)G'Q'.
\eqno (24)
$$
$\tilde{t\1}$ means the $2(n+1)\times 2(n+1)$ matrix
$$
\pmatrix{t&\ph&\ph&\ph&\ph&\ph&\ph&\ph\cr
         \ph&\ddots &\ph&\ph&\ph&\ph&0&\ph\cr
         \ph&\ph&t&\ph&\ph&\ph&\ph&\ph\cr
         \ph&\ph&\ph&1&\ph&\ph&\ph&\ph\cr
         \ph&\ph&\ph&\ph&1&\ph&\ph&\ph&\cr
         \ph&\ph&\ph&\ph&\ph&t&\ph&\ph\cr
         \ph&0&\ph&\ph&\ph&\ph&\ddots&\ph\cr
         \ph&\ph&\ph&\ph&\ph&\ph&\ph&t\cr}
\eqno (25)
$$
Now $G'=(H_0(t)+\eta \1)^{-1}$ is expanded to second order in $\eta$,
 ($G_0=H_0(t)^{-1}$),
$$
G'(t,\eta)\simeq G_0(t)(\1 -\eta G_0(t)+\eta ^2G_0^2(t)).
\eqno (26)
$$
Therefore, the determinant of (24) becomes
$$
\det(C+(1-t)G_0(-\eta G_0+\eta ^2G_0^2)Q')
\eqno (27)
$$
with
$$
C=\tilde{t\1}+(1-t)G_0Q'
\eqno (28)
$$
Eventually, the determinant factorizes
$$
{\overline{S(n)}\ }^2=
\det C\det (\1 +(1-t)C^{-1}G_0(-\eta G_0+\eta ^2 G_0^2)Q')
\eqno (29)
$$
The matrix $C$ gives the averaged correlation function if $\eta$ would be
absent. Because we are in the region where the effective coupling $t$ is
greater than $t_c$, $\det C$ reaches a constant nonzero value in the limit
$n\rightarrow\infty$:
$$
\lim_{n\rightarrow\infty}\det C=[1-(\sinh 2\b)^{-4}]^{1/2}
\eqno (30).
$$
$\b = atanh (t)$.
Equation (30) gives the averaged square
of the magnetization in the ``outer''
phase, i.e. for $t<t_-$ or $t>t_+$. It leads to a nonvanishing magnetization
for $t>t_+$ as in the pure system, albeit the functional dependence of the
effective $t$ on $t_0$ has to be taken into account.
However,in the framework of the $N\rightarrow \infty$ limit
the averaged magnetization exhibits a finite jump at the point $t_+$.
That means, the phase transition from the magnetically ordered to the
disordered state occurs at a temperature corresponding to $t_+$. The
spontaneous magnetization vanishes for $t<t_+$, whereas it is different from
zero for $t>t_+$.
To see this,
we calculate the second determinant in (29) for $0<\eta \ll 1$.

\no
$C$ has the following block structure$^1$
$$
\pmatrix{\ph&\ph&\ph&0&\ph&\ph&\ph&\ph\cr
         \ph&A&\ph&\vdots&\ph&\ph&0&\ph\cr
         \ph&\ph&\ph&0&\ph&\ph&\ph&\ph\cr
         b_0&\ldots&b_{n-1}&1&\ph&\ph&\ph&\ph\cr
         \ph&\ph&\ph&\ph&1&b_{n-1}^*&\ldots&b_0^*\cr
         \ph&\ph&\ph&\ph&0&\ph&\ph&\ph\cr
         \ph&0&\ph&\ph&\vdots&\ph&A^{\dag}&\ph\cr
         \ph&\ph&\ph&\ph&0&\ph&\ph&\ph}
\eqno
(31)
$$
$A$ is a $n\times n$ Toeplitz matrix with the elements
$$
A_{j,k}=\int_{-\pi}^{\pi}{dp\over {2\pi}}f(p)e^{ip(j-k)}
\eqno (32)
$$
where we choose the indices $j,k$ to run from $-(n-1)/2$ to $(n-1)/2$,
and
$$
b_j=\int_{-\pi}^{\pi}{dp\over{2\pi}}f(p)e^{ip(n-j)}
\eqno (33)
$$
for $j\in [0,n-1]$.
$f(p)$ is an unimodular function,
parametrized by $t$ and its dual $\tilde{t}$,
$\tilde{t}=(1-t)/(1+t)$:
$$
f(p)={\Big(}{{(t-\tilde{t}e^{ip})(t\tilde{t}-e^{ip})}
\over{(te^{ip}-\tilde{t})(t\tilde{t}e^{ip}-1)}}{\Big)}^{1/2}.
\eqno (34)
$$
The second factor in equation (28) can now be written in the form
$$
\exp (Tr \log (\1 + (1-t)C^{-1}G_0(-\eta G_0 +\eta ^2G_0^2)Q'))
\eqno (35)
$$
The logarithm can be expanded into
powers of $\eta$, because we are not in the
critical regime and therefore $G_0(t)$ is a bounded operator in its
domain of definition and has finite operator norm. By an appropriate choice
of $\eta$, (close enough to zero), the expansion is welldefined.
In first approximation higher terms then quadratic are neglected.
 In the appendix
it is shown that the term proportional to $\eta$
$$
Tr (1-t)C^{-1}G_0^2Q'
\eqno (36)
$$
vanishes, and the term proportional to
$\eta^2$ is negative definite, yielding
finally for large $n$
$$
{\overline{S(n)}\ }^2\simeq const.\times \exp [-(\eta^2\gamma (t))n]
\eqno (37)
$$
where $\gamma(t)$ is a positive
constant, depending on $t_0$ and the disorder strenght
$g$. One sees that the disorder averaged spin correlation function decays
exponentially for an arbitrary nonzero value of $\eta$. Eventually, the
averaged spontaneous magnetization vanishes in the phase with $\eta \neq 0$.

\no
{\bf Conclusions}

\no
In this article, the phase transition from the ferromagnetic ordered
phase to the disorder dominated phase in the two dimensional Random
Bond Ising Model is investigated. Whereas in previous approaches$^{2,3,4}$
a large scale approximation in the vicinity of an assumed critical region
allowed to reduce the number of independent fermion degrees of freedom from
four to two, we treat the full lattice theory in a selfconsistent way in the
framework of a $1/N$ expansion. This expansion is not supposed to correspond
to some special physical properties
of the system, but serves as a bookkeeping
device, which allows to extract the most
important ``mean field'' contributions
usually omitted in renormalization
group approaches (i.e. tadpoles). The most
general selfconsistent ansatz for the effective Hamiltonian has the same
structure as the saddle point solution of the Q-matrix theory$^3$, although
the reason for introducing a possibly nonvanishing $\eta$ is different: In
the Q-matrix theory one uses supersymmetric fields and the $\eta$-term
regularizes the integration over the bosons. After averaging over the
disorder,
$\eta$ becomes nonzero in a certain region due to spontaneous symmetry
breaking. The order parameter in the symmetry broken phase can therefore be
identified with $\eta$, leading to a continous phase transition from the
ferromagnetic to the disorder dominated phase as well as from this phase
to the paramagnetic one. The question arises, whether the spontanous
magnetization in the new phase vanishes or not. To this end, we calculated
the asymptotic, ($n\rightarrow\infty$), behavior of the disorder
averaged spin correlation function in a region around the transition line
from the magnetic ordered phase to the new phase. Whereas the magnetization
on the ferromagnetic side is finite
corresponding to $T<T_c$, the correlation
function decays exponentially
$$
\overline{S(n)}\simeq \exp -n/\xi
$$
with
$$
\xi={2\over{\g(t)\eta^2}}.
$$
That means, $\eta^2$ plays the role of an inverse correlation length.
However, it is not clear what kind of operators becomes massless as $\xi$
goes to infinity. The Q-matrix analysis shows a massless mode of the
local composite Q-operators at $t_+$.
Eventually, a rigorous RG-treatment of the
transition point has to deal with these objects.
The temperature dependence of the magnetization in the ferromagnetic phase
can be calculated by means of equations (18) and (30).
 The appearance of $\eta^2$ and not $\eta$ in the expression for
the averaged correlation function is
due to the fact, that the selfconsistent
equation allows two real solutions with $\eta$ positive or negative, whereas
the physics must not depend on the sign of $\eta$. This corresponds to the
freedom to choose the sign of the regularization term in the supersymmetric
theory. If one relates the averaged asymptotic correlation function to the
averaged square of the spontaneous
magnetization in the usual way, it follows,
that the new phase shows no long
range magnetic order. Therefore, the
only possibility to discern it from the
paramagnetic region consists in the
investigation of the relaxational dynamics
for  the spins, which will be discussed elsewhere. However, the present
article says nothing about the transition from the new phase to the
paramagnetic one. It is even possible, that something
happens in the new phase itself,
when the effective $t$ reaches $t_c$. In the framework of
the $N\rightarrow\infty$, resp. saddlepoint calculation, the spontaneous
magnetization exhibits a finite jump at the transition
from the ordered to the new phase. Whether this remains correct if one takes
into account fluctuations can perhaps
be answered via a renormalization group
treatment of the theory on the ferromagnetic side.

\no
{\bf {Aknowledgements.}}

\no
I may thank K.Ziegler for many
important suggestions. This work was supported
by a grant from the ``Graduiertenf\"orderung
des Landes Baden-W\"urttemberg''.

\vfill\eject
\no
{\bf Appendix}

\no
To calculate the term proportional to $\eta$ in equation (36), we write
$G_0(\p)$ as
$$
G_0={1\over{\det H_0}}\pmatrix{\a_1&\b_1&\e&\d\cr
                               -\b_1^*&\a_1^*&-\d^*&\e^*\cr
                               -\e^*&\d&\a_2&\b_2\cr
                               -\d^*&-\e&-\b_2^*&\a_2^*}
\eqno (A1)
$$
with
$$
\a_1=b-b^*
$$
$$
\a_2=a^*-a
$$
$$
\b_1=(b+b^*)-abb^*
$$
$$
\b_2=(a+a^*)-baa^*
\eqno (A2)
$$
$$
\e=ab^*-2
$$
$$
\d=ab-2
$$
with the conventions of equation (11). The (3,4)- (i.e. horizontal)
sector of $G_0^2(\p)$ reads then
$$
{1\over{\det H_0^2}}\pmatrix{-(|\a_2|^2+|\b_2|^2+|\e|^2+|\d|^2)&0\cr
           0&-(|\a_2|^2+|\b_2|^2+|\e|^2+|\d|^2)}
\eqno (A3)
$$
Therefore, $G_0^2Q'$ has the form
$$
\pmatrix{{\bf 0}&{\bf *}\cr
         {\bf *}&{\bf 0}}
\eqno (A4)
$$
where $\bf 0$ and $\bf *$ are $(n+1)\times (n+1)$ matrices.
multiplying from the left with $C^{-1}$ yields a matrix with vanishing
diagonal entries. It follows that the contribution proportional to $\eta$
vanishes.

\no
To calculate the quadratic term, we write
$$
Tr(C^{-1}G_0^{3}Q')=Tr(G_0^2(C-\tilde{t\1})C^{-1})=\tilde{Tr}(G_0^2)-
\tilde{Tr}(G_0^2
\tilde{t\1}C^{-1})
\eqno (A5)
$$
and $\tilde {Tr}$ means the trace in the $2(n+1)\times 2(n+1)$ dimensional
subspace of the horizontal sector under consideration.
To give an estimate for the matrix $C^{-1}$, we use the fact that the
matrix $A$ in equation (31) is ``almost unitary'', in the sense that
$$
AA^{\dag}=\1+{1\over n}B
\eqno (A6)
$$
where $B$ has bounded matrix elements. This is due to the fact that
$$
\sum_{j=-(n-1)/2}^{(n-1)/2}g(p)e^{ipj}=g(0)+O(1/n)
\eqno (A7)
$$
for a bounded function $g(p)$, and $f(p)$ in equation (32) is of unit
modulus. Then it is possible to write for $C^{-1}$
$$
C^{-1}=\pmatrix{\ph&\ph&\ph&0&\ph&\ph&\ph&\ph\cr
         \ph&A^{\dag}&\ph&\vdots&\ph&\ph&0&\ph\cr
         \ph&\ph&\ph&0&\ph&\ph&\ph&\ph\cr
         d_0&\ldots&d_{n-1}&1&\ph&\ph&\ph&\ph\cr
         \ph&\ph&\ph&\ph&1&d_{n-1}^*&\ldots&d_0^*\cr
         \ph&\ph&\ph&\ph&0&\ph&\ph&\ph\cr
         \ph&0&\ph&\ph&\vdots&\ph&A&\ph\cr
         \ph&\ph&\ph&\ph&0&\ph&\ph&\ph}
+{1\over n}B'
\eqno (A8)
$$
with
$$
d_j=-\sum_{i=0}^{n-1}b_iA^{\dag}_{i-(n-1)/2,j-(n-1)/2}=O(1/n)
\eqno (A9)
$$
With the definition
$$
G_{0,jk}^2=G_0^2(x_1=j,y_1=0;x_2=k,y_2=0)=\int_{-\pi}^{\pi}{dp\over{2\pi}}
h(p)e^{ip(j-k)}
\eqno (A10)
$$
for $j,k\in [0,n]$, we get
$$
\tilde{Tr}(G_0^2\tilde{t\1}C^{-1})=2G_{0,nn}^2+
2nti\int_{-\pi}^{\pi}{dp\over{2\pi}}
h(p)Im(f(p)) +O(1)
\eqno (A11)
$$
The contribution from the integral
 leads to at most an oscillating factor for the
averaged spin correlation function, but actually it vanishes due to the
symmetry properties of $h(p)$ and $f(p)$. It follows that the relevant
contribution is the first term in equation (A5), namely
$$
\tilde{Tr}G_0^2=2(n+1)G_{0,nn}.
\eqno (A12)
$$
Finally one gets for the (positive definite) $\g(t)$ of equation (37)
$$
\g(t)=2(1-t)\int_{-\pi}^{\pi}{{dp_xdp_y}\over{(2\pi)^2}}
{{|\a_2|^2+|\b_2|^2+|\e|^2+|\d|^2}\over{\det H_0(t)^2}}
\eqno (A13)
$$
\vfill\eject
\no
{\bf References:}

\no
[1] C.Itzykson, J.M.Drouffe
``Statistical field theory'', Part One, Cambridge
University Press, 1989

\no
[2] D.Braak and K.Ziegler Z.Phys B, {\bf 89}, 1992, 361

\no
[3] K.Ziegler Nucl.Phys.{\bf B344} (1990) 499-530

and Europhys.Lett. {\bf 14} (5) (1991) 451-420

\no
[4] V.Dotsenko and Vl.S.Dotsenko Adv.Phys.{\bf 32} (1983) 129

\no
[5] B.M.McCoy and T.T.Wu Phys.Rev. B {\bf 2} (1970) 2795

\no
[6] R.Shankar Phys.Rev.Lett. {\bf 58} (1987) 2466

\no
[7] B.N.Shalaev Sov.Phys.Solid State {\bf 26} (1984) 1811

\no
[8] A.W.W.Ludwig Nucl.Phys.{\bf B285} (1987) 97

and Phys.Rev.Lett. {\bf 61} (1988) 2388

\no
[9] A.W.W.Ludwig Nucl.Phys.{\bf B330} (1990) 639

\no
[10] P.N.Timonin Zh.Eksp.Teor.Fiz. {\bf 95} (1989) 893

\no
[11] K.Ziegler J.Phys. A {\bf 19} (1986) L943

\no
[12] K.Ziegler J.Phys. A {\bf 21} (1988) L661

\no
[13] D.Braak (unpublished)

\no
[14] K.Ziegler Nucl.Phys.{\bf B280} (1987) 661

\no
[15] K.Ziegler Nucl.Phys.{\bf B285} (1987) 606

\no
[16] D.Braak Bonn-IR--90-37 (1990) (unpublished)

\no
[17] J.S.Wang, W.Selke, Vl.S.Dotsenko and V.B.Andreichenko Physica A164 (2)
221-239

\no
[18] A.L.Talapov, V.B.Andreichenko, Vl.S.Dotsenko and L.N.Shchur

\no
in ``Computer simulations in condensed matter physics'' eds. D.P.Landau,
K.K.Mon and H.B.Schuettler (Springer, Heidelberg, 1991)

\no
[19] M.F\"ahnle, T.Holey and
J.Eckert J.Magn.Magn.Mater. {\bf{104-107}} (1992)
195-196

\no
[20] J.P.Jamet, C.Landee, J.Ferr\'e,
M.Ayadi, H,Gaubi, I.Yamada and K.Ziegler,
preprint Orsay (1992)

\no
[21] J.Glimm and A.Jaffe ``Quantum Physics'', 2.ed., Springer-Verlag 1987

\no
[22] K.Binder and A.P.Young Rev.Mod.Phys {\bf{58}} (1986) 801
\vfill\eject
\no
{\bf Figure Captions:}

\no
Fig. 1: Two off diagonal contributions to the propagator, coming from
averaging over bond disorder in y-direction. The left graph corresponds
to equation (6). $\a$ and $\b$ are indices of the $N$ colors. $\a$ is fixed
whereas $\b$ runs from $1$ to $N$.

\no
Fig. 2: Two diagonal contributions. These vanish if $\eta =0$.

\no
Fig. 3: A two-propagator term in the expansion
of $\langle\exp {1\over2}\xi Q\xi\rangle_{Z_D}$.
The horizontal line indicates the section
of the x-axis between the two spins.
The distance between them in units of the lattice spacing is $n$.

\no
Fig. 4: A term in the average of the
expression $\langle\exp {1\over2}\xi Q\xi\rangle_{Z_D}$,
taken into account using the effective $G'$ from section 1.
$\a$ is a fixed replica index and $\b$ runs from 1 to $N$.

\no
Fig. 5: A term taken into account by replacing $t_0$ by $t$. $\b$ and $\g$
run from $1$ to $N$.

\no
Fig. 6: A term which is omitted in the $N\rightarrow\infty$ approximation.

\no
Fig. 7: The effective bond strenght $t$ as function of $t_0$. The vertical
lines denote $t_-=0.18$, resp. $t_+=0.71$, the
new phase lies between them. The value
of $g$ is 0.3.
\vfill\eject\bye